\documentclass{article}
\newtheorem{theorem}{Theorem}
\usepackage{graphicx}
\RequirePackage{fix-cm}
\usepackage{lipsum}
\usepackage{textcomp}
\usepackage{amsmath}
\usepackage{mathtools}
\usepackage{arxiv}
\usepackage[utf8]{inputenc} 
\usepackage[T1]{fontenc}    
\usepackage[colorlinks = true,
           linkcolor = blue,
            urlcolor  = blue,
           citecolor = blue,
            anchorcolor = black]{hyperref}      
\usepackage{url}            
\usepackage{booktabs}       
\usepackage{amsfonts}       
\usepackage{nicefrac}       
\usepackage{microtype}      
\usepackage{lipsum}
\usepackage{multirow}
\title{Multi-objective scheduling on two dedicated processors}

\author{
  Adel KACEM\thanks{University of Sfax, Modeling and Optimization for Decisional, Industrial and Logistic Systems Laboratory. } \\
  \texttt{adel.kacem@gmail.com} \\
   \And
 Abdelaziz DAMMAK \thanks{University of Sfax, Faculty of Economics and Management, Airport Street, km 4, Post Office Box 1088, 3018 Sfax, Tunisia.}\\
  \texttt{abdelaziz.dammak@fsegs.rnu.tn} \\
}

\begin{document}
\maketitle

\begin{abstract}
 \textcolor{red}{}We study a multi-objective scheduling problem on two dedicated processors. The aim is to minimize simultaneously the makespan, the total tardiness and the total completion time. This NP-hard problem requires the use of well-adapted methods. For this, we adapted genetic algorithms to multi-objective case. Three methods are presented to solve this problem. The first is aggregative, the second is Pareto and the third is non-dominated sorting genetic algorithm II (NSGA-II). We proposed some adapted lower bounds for each criterion to evaluate the quality of the found results on a large set of instances. Indeed, these bounds also make it possible to determine the dominance of one algorithm over another based on the different results found by each of them. We used two metrics to measure the quality of the Pareto front: the hypervolume indicator (HV) and the number of solutions in the optimal front (ND). The obtained results show the effectiveness of the proposed algorithms.
\end{abstract}

\keywords{scheduling \and tasks\and dedicated processors\and total tardiness\and
makespan\and completion time\and lower bound\and NSGA-II\and Pareto front.}

\section{Introduction}
\label{intro}
This work aims at optimizing the computer control systems when these systems have two dedicated processors: the assignment of tasks to these processors is fixed. For this problem, we have three types of tasks. Some tasks must be processed only by the first processor, others by the second processor, and the remaining tasks need simultaneously both processors. This problem represents a practical issue in computer control systems, where a task is performed in several copies on different processors in order to ensure better safety of the system. In production management, we can cite the case where a task requires several operators for its execution.\\ 
The contribution of our work is to propose lower bounds for the three studied criteria (makespan, total tardiness and total completion time) and to develop genetic algorithms to solve this problem in the multi-objective case. The lower bounds allow us to assess the quality of the feasible solutions and the genetic algorithms incorporates the optimization part. We implemented our approach by considering aggregative, NSGA-II and Pareto scenarios on a large set of instances. The results show the effectiveness of the implemented algorithms.  The studied problem is $P2|fix_j,r_j|C_{max}, \sum T_j , \sum C_j $ according the standard ternary notation, where $P2$ denotes two processors; $fix_j$ indicates that each task has one or two dedicated processors and the assignment of each task is fixed; $r_j$ denotes the release date; $C_{max}$, $\sum T_j$ and $\sum C_j$ indicates the makespan, the total tardiness and the total completion time respectively.\\
The next section is a review of existing research related to the studied problem. In Section \ref{Problem}, a mathematical formulation model is proposed, some notations are detailed and the proposed lower bounds for the makespan, total completion time and total tardiness are given. In Section \ref{solv-app}, we present the solving approaches. Three methods to solve the considered problem are developed. The first one is aggregative with Uniform Design, the second is Pareto and the third is the NSGA-II. Section \ref{Numerical} deals with the generation of instances, the proposed metrics to measure the quality of the Pareto front, the computational results and the qualitative and quantitative analysis. Finally concluding remarks are given in Section \ref{conc}.
\section{Literature Review}
\label{review}
Few studies have dealt with this problem. The most important studies are mentioned in the following paragraphs.\\ 
Coffman et al. \cite{Coffman1985} studied the file transfer problem in the field of computer networks where each computer has a number of different ports for data exchange. File transfer uses a subset of ports, therefore a multiprocessor task on dedicated processors. The boot time of the transfers is also taken into account, then different transfer protocols are proposed, and performance results are demonstrated. Drozdowski \cite{Drozdowski1996} cited this paper to describe the actual applications of scheduling problems on dedicated processors.\\
Craig et al. \cite{Craig1988} studied the problem in testing integrated circuits VLSI (very large-scale integration). To test a component of these circuits, several other electronic components are needed simultaneously. The authors addressed the problems in case when the processing times are unitary or arbitrary. A heuristic based on the maximum degree of incompatibility has been proposed to solve these two problems \(P|fix_j,p_j=1|C_{max}\) and \(P|fix_j|C_{max}\) ($p_j$ denotes the processing time of task $j$).\\
These works (Coffman et al. \cite{Coffman1985} ; Craig et al. \cite{Craig1988}), which cover several fields of application, made it possible to form the first theoretical basis for scheduling problems on dedicated processors. This topic has been widely investigated during the past years. The most remarkable work has been devoted to the study of the complexity.\\
Hoogeveen et al. \cite{Hoogeveen1994} showed that the problem \(P2|fix_j|\sum w_j C_j\) is NP-hard in the strong sense ($w_j$ the weight and $C_j$ the completion time of task $j$). The preemption of tasks does not make the problem easier. Oguz and Ercan \cite{Oguz2005} proved that the problem  \(P2|fix_j, pmtn|\sum w_j C_j\) is NP-hard in the strong sense ($pmtn$ allows us the preemption of tasks). Afrati et al. \cite{Afrati1999} proposed a polynomial time approximation scheme (PTAS)  for the problem \(Pm|fix_j, pmtn|\sum C_j\) and a second PTAS approximation scheme proposed by Afrati and Milis \cite{Afrati2006}.\\
Chu \cite{Chu1992} proposed a lower bound for the minimization of total tardiness problem; the calculation involves the SRPT priority rule (Shortest Remaining processing Time) for a relaxed problem with preemption. The main idea is that each time the processor becomes available, an unfinished task available with the shortest remaining processing time is set. The execution of a task is interrupted when its remaining processing time is strictly greater than the length of processing task that becomes available.\\
Leung and Wang \cite{Leung2000} proposed a genetic algorithm with multiple fitness functions to conduct research in order to solve a multi-objective problem. The authors applied an experimental design method called Uniform Design to select the weights used with the objective functions and diversify uniformly selected solutions.\\
Kacem \cite{Kacem2007} developed two lower bounds for tardiness minimization problem on a single machine with Family Setup Times. The first lower bound is based on Emmons theorem \cite{Emmons1969} and the SPT rule (Shortest Processing Time), the second is achieved by sorting tasks by processing times and the idea of due dates exchange. Another idea of solving the linear programming problem, was also proposed.\\
Berrichi et al. \cite{Berrichi2007} studied a bi-objective model of parallel machine problem using reliability models to take into account the service side. Two genetic algorithms were developed to obtain an approximation of the Pareto front: One algorithm that uses the two objectives weighted and NSGA-II algorithm.\\
Rebai et al. \cite{Rebai2010} introduced three lower bounds for minimization tardiness problem on one machine to schedule preventive maintenance tasks. The first lower bound is based on the Lagrangian relaxation of mathematical model. The second is obtained by the sum of M costs calculated for M tasks, and the third is an adaptation of the lower bound given by Li\cite{Li1997} for the problem of earliness tardiness minimization with a single due date for each task.\\
Manaa and Chu \cite{Manaa2010} proposed a branch-and-bound method to minimize the makespan. In their article, the authors presented a lower bound that has been proven. This method can treat all instances generated up to 30 tasks for the most difficult cases in less than 15 minutes.\\
Vallada and Ruiz \cite{Vallada2011} studied the unrelated parallel machine scheduling problem. A genetic algorithm is developed to solve this problem. The proposed method includes a fast local search and a local search enhanced crossover operator.  The computational and statistical analysis shows an excellent performance in a comprehensive benchmark set of instances.\\
Bradstreet \cite{Bradstreet2011} introduced the hypervolume indicator (HV) to measure the quality of the Pareto front. The hypervolume is one of the most famous indicator that can reflect the dominance of Pareto fronts.\\
Alhadi et al. \cite{Alhadi2019} proposed an approximation algorithms to minimize the maximum lateness and makespan on parallel machines. In this paper, the authors presented polynomial time approximation schemes to generate an approximate Pareto Frontier.\\
Kacem and Dammak \cite{Kacem2019} studied the problem of bi-objective scheduling of multi-processor tasks on two dedicated processors. The authors adapted the genetic algorithm to solve the problem of minimizing the makespan and the total tardiness for the large size instances. The results found showed the effectiveness of the proposed genetic algorithms and the encouraging quality of the lower bounds constructed in (Manaa and Chu \cite{Manaa2010}; Kacem and Dammak \cite{Kacem2017}). For that, we decided to study a new extension of this problem by introducing the total completion time criterion. According the standard ternary notation, the studied problem is $P2|fix_j,r_j|C_{max}, \sum T_j , \sum C_j $.
\section{Problem statement}
\label{Problem}
In this section, we detail some notations, we propose a mathematical model for our studied problem and we give a lower bounds for the makespan, total completion time and total tardiness criteria. 
\subsection{Notation}
\label{sec:31}
The following fields denote: 
\begin{itemize}
\item $P2$: Two processors.
\end{itemize}
\begin{itemize}
\item $fix_j$: Each task has one or two dedicated processors and the assignment of each task is fixed.
\end{itemize}
\begin{itemize}
\item $r_j$: Release date of task $j$.
\item $C_{max}$: Makespan.
\end{itemize}
\begin{itemize}
\item $d_j$: Due date of task $j$.
\end{itemize}
\begin{itemize}
\item $T_j$: Tardiness of task $j$; $T_j=max\left\{ C_j-d_j;0\right\}$with $C_j$ the completion time of task $j$.
\end{itemize}
\begin{itemize}
\item $p_j$: Processing time of task $j$.
\end{itemize}
\subsection{Mathematical model}
\label{sec:32}
We propose here a mathematical formulation of our multi-objective scheduling problem. We give a set of parameters and variables necessary for our model.
\[x_{l,k} = \left\{ 
\begin{array}{l l}
  1 & \quad \text{if job $l$ completes before job $k$ starts}\\
  0 & \quad \text{else;}\\ \end{array} \right. \]
$P_1$ is a set of jobs requiring the first processor.\\
$P_2$ is a set of jobs requiring the second processor.\\
$P_{1,2}$ is a set of jobs requiring both processors simultaneously.\\
$M$ is a constant penalty. 
\begin{equation}
\text{Minimize }\{ C_{max}, \sum T_j , \sum C_j\}
\end{equation}
Subject to:\\
\begin{equation}
C_k\geq C_l+p_k+(x_{l,k}-1).M \text{  } \forall(l,k) \in (P_1\cup P_{1,2})^2 \text{ with }l\neq k
\end{equation}
\begin{equation}
x_{l,k}+x_{k,l}=1 \text{  } \forall(l,k) \in (P_1\cup P_{1,2})^2 \text{ with }l\neq k
\end{equation}
\begin{equation}
C_k\geq C_l+p_k+(x_{l,k}-1).M \text{  } \forall(l,k) \in (P_2\cup P_{1,2})^2 \text{ with }l\neq k
\end{equation}
\begin{equation}
x_{l,k}+x_{k,l}=1 \text{  } \forall(l,k) \in (P_2\cup P_{1,2})^2 \text{ with }l\neq k
\end{equation}
\begin{equation}
C_{max}\geq C_j \text{  } \forall j \in (P_1 \cup P_2\cup P_{1,2})
\end{equation}
\begin{equation}
T_j\geq C_j-d_j \text{  } \forall j \in (P_1 \cup P_2\cup P_{1,2})
\end{equation}
\begin{equation}
T_j\geq 0 \text{  } \forall j \in (P_1 \cup P_2\cup P_{1,2})
\end{equation}
\begin{equation}
C_j\geq r_j+p_j \text{  } \forall j \in (P_1 \cup P_2\cup P_{1,2})
\end{equation}
\begin{equation}
x_{k,l} \in \{0,1\} \text{  }\forall k\neq l \text{  }\forall(k,l) \in ((P_1\cup P_{1,2})^2 \cup (P_2\cup P_{1,2})^2)
\end{equation}
Equation (1) expresses the three criteria: the makespan, the total tardiness and the total completion time. In the constraints (2) and (4), for each job $k$ sequenced after job $l$ completes, the completion time of job $k$ is greater than or equal to the completion time of job $l$ plus the processing time of job $k$. The constraints (3) and (5) make it possible two jobs to complete one before the second job starts. The constraint (6) ensures that the makespan is greater than or equal to the completion time for each job $j$. The constraints (7)-(9) compute the tardiness. The completion time of job $j$ is greater than or equal to the release date plus the processing time of job $j$.
The constraint (10) defines the domain of definition of the parameters of the model.\\
We study three scheduling problems (makespan, total tardiness and total completion time) on two dedicated processors. To assess the quality of the results found by such a method, we use the following lower bounds.
\subsection{Lower bound $LBC$ for problem \(P2|fix_j,r_j|C_{max}\)}
\label{sec:33}
Manaa and Chu \cite{Manaa2010} proposed two ideas to construct a lower bound for the considered problem:\\
\begin{itemize}
\item The idea of dividing the problem into two sub-problems on one processor by relaxing the studied problem.
\end{itemize}
\begin{itemize}
\item The idea of Bianco et al. \cite{Bianco1997} an optimal solution to minimize the makespan for one-processor problem.
\end{itemize}
The relaxation of the studied problem allows us to obtain two simple problems:\\
a)	Scheduling tasks that necessitate using simultaneously both processors and tasks that require the first processor.\\
b)	Scheduling tasks that require employing simultaneously both processors and tasks that necessitate the second processor.\\  
The optimal solutions of problems (a) and (b) can be found by scheduling tasks according to the order of their release dates.\\
The lower bound  for the studied problem corresponds to the maximum value of the solutions of problems (a) and (b).
\subsection{Lower bound $LBTC$ for problem \(P2|fix_j,r_j|\sum C_j\)}
\label{sec:34}
\begin{figure}
\centering
\includegraphics[width=12 cm,height=1.5 cm]{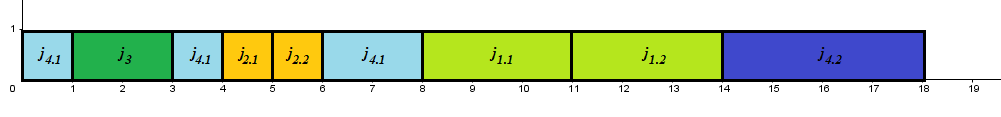}
\caption{ SRPT representation on P1}
\label{fig:1}      
\end{figure}

In this study, we use and combine three ideas to build a lower bound: 
\begin{itemize}
\item The idea of reducing the problem into two sub-problems on one processor by partitioning the bi-processor tasks. 
\item The idea of dividing the mono- processor tasks into two tasks.
\item The idea of under-estimating the completion times of the tasks on a single processor (originally proposed by Chu \cite{Chu1992}).
\end{itemize}
The first of this lower bound is to partition the bi- processor tasks into two mono-processor tasks, each of them on one of the processors. We get two independent problems on each processor.\\
On the first processor$P_1$, we consider the $n_1$ mono-processor tasks $j_j$ with a weight $w_j^1=1$, and the  $n_{12}$ bi- processor sub-tasks $j_j$ on processor $P_1$ a weight $w_j^1=w$, with $w \in \left [0,1 \right]$.\\
Similarly, we consider the second processor $P_2$ , the $n_2$ mono-processor tasks $j_j$ with a weight $w_j^2=1$. However, the bi-processor sub-tasks  $j_j$ on processor $P_2$ a weight $w_j^2=1-w$. Thus, we obtain a problem on each processor:
\(P_1|fix_j,r_j|\sum w_j^1 C_j\) and \(P_2|fix_j,r_j|\sum w_j^2 C_j\).\\
We consider $w=\frac{1}{2}$. The next step is to divide the mono-processor tasks (with a weight  $w_j^1=1$) in two tasks. We get for each divided task two sub tasks $j_{j1}$ and  $j_{j2}$ with release date $r_{j1}=r_j$ ; $r_{j2}=r_j+\frac{p_j}{2}$ and processing time $p_{j1}=p_{j2}=\frac{p_j}{2}$.\\
We divide the weight on for each sub tasks. We are getting $w_j^1=\frac{1}{2}$ if  $\forall j_j \in \left\{P_1;P_{12}\right\}$. From where, \\$Lb1=Lb\left(P_1|fix_j,r_j|\sum w_j^1 C_j\right) =Lb\left(P_1|fix_j,r_j|\sum \frac{1}{2}  C_j\right) +\sum_{\\j\in P_{1}}^{} {\frac{p_j}{4}}$, with  $\sum_{\\j\in P_{1}}^{} {\frac{p_j}{4}}$ is a penalty to be added according to Webster formula.\\
$Lb1=Lb\left(P_1|fix_j,r_j|\sum \frac{1}{2}  C_j\right) +\sum_{\\j\in P_{1}}^{} {\frac{p_j}{4}}=\frac{1}{2} Lb\left(P_1|fix_j,r_j|\sum C_j\right) +\sum_{\\j\in P_{1}}^{} {\frac{p_j}{4}}$.\\
We apply the same principle for the problem \(P_2|fix_j,r_j|\sum w_j^2 C_j\). We are getting $Lb2=Lb\left(P_2|fix_j,r_j|\sum \frac{1}{2}  C_j\right) +\sum_{\\j\in P_{2}}^{} {\frac{p_j}{4}}=\frac{1}{2} Lb\left(P_2|fix_j,r_j|\sum C_j\right) +\sum_{\\j\in P_{2}}^{} {\frac{p_j}{4}}$.\\
$LBTC=Lb1+Lb2$ is then a lower bound for problem \(P2|fix_j,r_j|\sum C_j\). The calculation of the lower bounds of completion times for the problem on each
processor uses the following theorem for  \(1|r_j, pre|\sum C_j\). 
\begin{theorem} (Chu \cite{Chu1992})
\\Let $C_{\left[i\right]}\left (\sigma\right) $ be the completion time of the task in the $i^{th}$ position of a feasible schedule .
$C'_i$ is the completion time of the task in the $i^{th}$ position of a feasible schedule constructed by the SRPT (Shortest Remaining Processing Time) priority rule.
Chu proved that for every feasible schedule $\sigma$, we have: \(C_{\left[i\right]}\left (\sigma\right)\geq C'_i\)
\end{theorem}
By applying the theorem (Chu \cite{Chu1992}), we compute a lower bound on the completion time of each job.\\
Example : Let us consider the instance in Table \ref{tab:1}.\\
\begin{table}
\caption{Example}
\centerline{
\label{tab:1}     
\begin{tabular}{p{2.5cm} p{2.5cm} p{2.5cm} p{2.5cm}}
\hline\noalign{\smallskip}
$j$ & $r_j$ & $p_j$  & $P$ \\
\noalign{\smallskip}\hline\noalign{\smallskip}
$j_1$ & $2$ & $6$ & $P_1$ \\
$j_2$ & $4$ & $2$ & $P_1$ \\
$j_3$ & $1$ & $2$ & $P_{12}$ \\
$j_4$ & $0$ & $8$ & $P_1$ \\
$j_5$ & $3$ & $2$ & $P_2$ \\
$j_6$ & $2$ & $6$ & $P_2$ \\
$j_7$ & $1$ & $2$ & $P_2$ \\
\noalign{\smallskip}\hline
\end{tabular}}
\end{table}
We apply the principle of calculation of the lower bound mentioned above and we get two sub-problems on each processor. On the first processor $P_1$, we schedule the tasks $\left\{ j_1,j_2,j_3,j_4\right\}$. We divide the tasks  $\left\{ j_1,j_2,j_4\right\} $ on into two, we get the following tasks: $\left\{ j_{11},j_{12},j_{21},j_{22},j_{41},j_{42}\right\} $with the following parameters described in Table \ref{tab:2}.\\
\begin{table}
\caption{Results division on $P_1$}
\centerline{
\label{tab:2}       
\begin{tabular}{p{2.5cm} p{2.5cm} p{2.5cm} p{2.5cm}}
\hline\noalign{\smallskip}
$j$ & $r_j$ & $p_j$  & $P$ \\
\noalign{\smallskip}\hline\noalign{\smallskip}
$j_{1.1}$ & $2$ & $3$ & $P_1$ \\
$j_{1.2}$ & $5$ & $3$ & $P_1$ \\
$j_{2.1}$ & $4$ & $1$ & $P_1$ \\
$j_{2.2}$& $5$ & $1$ & $P_1$ \\
$j_3$ & $1$ & $2$ & $P_1$ \\
$j_{4.1}$ & $0$ & $4$ & $P_1$ \\
$j_{4.2}$ & $4$ & $4$ & $P_1$ \\
\noalign{\smallskip}\hline
\end{tabular}}
\end{table}
The sequence built by the SRPT rule with preemption gives the solution described in Figure \ref{fig:1}.
\\The total of completion time, giving the following lower bound:\\ $Lb1=\frac{1}{2} Lb\left(P_1|fix_j,r_j|\sum C_j\right) +\sum_{\\j\in P_{1}}^{} {\frac{p_j}{4}}=\frac{3+5+6+8+11+14+18}{2}+\frac{3+3+1+1+4+4}{4}=36,5$. 
Respectively, we calculate the lower bound $Lb2$ for the problem $P_2|fix_j,r_j|\sum C_j$. Thus, we consider $LBTC=Lb1+Lb2$ as a lower bound for the problem $P2|fix_j,r_j|\sum C_j$.\\ .
\subsection{Lower bound $LBTT$ for the problem \(P2|fix_j,r_j|\sum T_j\)}
\label{sec:24}
Kacem and Dammak \cite{Kacem2017} proposed an adapted lower bound for the problem of minimization of total tardiness on two dedicated processors. The authors exploited and combined three ideas to construct this lower bound:\\
\begin{itemize}
\item The idea of reducing the problem in two sub-problems on one processor by partitioning the bi-processor tasks.
\end{itemize}
\begin{itemize}
\item The idea of under-estimating the completion time of the tasks (initially suggested by Chu \cite{Chu1992}).
\end{itemize}
\begin{itemize}
\item The idea of calculating the lower bound by assigning the due dates to the reduced completion times (originally proposed in Rebai et al. \cite{Rebai2010} for another scheduling problem).
\end{itemize}
The first step of this lower bound is to divide the bi-processor tasks into two mono-processor tasks; each of which is executed on one of the two processors. Consequently, we obtain two independent problems on each processor. On the first processor $P_1$, we consider the $n_1$ mono-processor tasks $J_j$ with a weight $\lambda_j^1=1$, and the $n_{12}$ bi-processor sub-tasks $J_j$ on processor $P_1$ having a weight $\lambda_j^1=\lambda$ with $\lambda \in \left [0,1 \right]$.\\
Similarly, we consider, on the second processor $P_2$, the $n_2$ mono-processor tasks $J_j$ with a weight $\lambda_j^2=1$ . However, the $n_{12}$ bi-processor sub-tasks $J_j$ on the processor $P_2$ have a weight $\lambda_j^2=1-\lambda$. Thus, we obtain a problem on each processor
\(P_1|fix_j,r_j|\sum\lambda_j^1 T_j\) and \(P_2|fix_j,r_j|\sum\lambda_j^2 T_j\).\\
Using the idea of Chu (described in the previous section), we compute a lower bound on the completion time of each task.\\
The next step of computing the lower bound is based on the idea of assigning the weight and the due date of each task to completion times' lower bounds. The total tardiness is minimized by the Hungarian algorithm.\\
Let $Cost_{i,j}$ be the cost of assigning a reduced $C'_i$ to the task $J_j$ supposed to end at the $i^{th}$ position of the schedule. This cost can be calculated according to the following formula: \(Cost_{i,j}=\lambda_j*max\left\{ C'_i-d_j;0\right\}\).\\
This assignment technique, presented by Rebai et al. \cite{Rebai2010}, allows us to elaborate a new lower bound. We apply the Hungarian algorithm to determine, from the assignment matrix $Cost_{i,j}^{P_1}$, a lower bound ($Lb1$) to solve the following problem \(P_1|fix_j,r_j|\sum\lambda_j^1 T_j\).\\
\begin{equation}
Lb1=min\sum x_{i,j}Cost_{i,j}
\end{equation}
\[\left \{
   \begin{array}{r c l}
     $$\sum_{i} x_{i,j}=1$$\\
     $$\sum_{j} x_{i,j}=1$$\\
     $$x_{i,j} \in \left\{0,1\right\}$$
   \end{array}
   \right. \]
Applying the same process, we calculate $Lb2$ for the  $P_2|fix_j,r_j|\sum\lambda_j^2 T_j$ problem. Thus, we consider $LBTT=Lb1+Lb2$ as a lower bound for the problem $P2|fix_j,r_j|\sum T_j$.\\
\textbf{Optimization of the lower bound $LBTT$} \\
To improve the constructed lower bound, we look for the weights $\lambda_j^*$, which maximize $LBTT$. The idea is to associate the better weight $\lambda_j^*$ for each bi-processor task $J_j$ which maximizes the tardiness calculated by the Hungarian algorithm. We use the following method to optimize the bound $LBTT$.
\begin{itemize}
\item For a bi-processor task $J_j \in (P_{1,2})$, we calculate the gap $e_j*$ between $T_{j^*}^{P_{1}}$ and $T_{j^*}^{P_{2}}$ (where $T_{j^*}^{P_{1}}$ (respectively $T_{j^*}^{P_{2}}$) is the tardiness of task $j$ associated to the sub-problem on $P_1$ (respectively $P_2$) obtained by the Hungarian algorithm).
\end{itemize}
\begin{itemize}
\item According to the gap, $e_j*$, we increase the $\lambda$ value for a negative gap (and we reduce it  respectively for a positive gap).
\end{itemize}
\begin{itemize}
\item We apply the Hungarian algorithm to the new matrix and we calculate a new lower bound $LBTT^\lambda$.
\end{itemize}
\begin{itemize}
\item We repeat this procedure $\forall J_j \in (P_{1,2})$.
\end{itemize}
Next, we present the study of the problem $P_2|fix_j,r_j|C_{max},\sum C_j, \sum T_j$.
\section{Solving approaches}
\label{solv-app}
We adapt a genetic algorithm to the multi-objective case. We propose three methods to solve the considered problem. The first is aggregative, second is Pareto and the third is the NSGA-II.\\
\subsection{The genetic algorithm}
\label{sec:41}
To represent the data of the studied problem, we used a standard coding technique. This coding consists in representing an individual with a permutation containing $N$ distinct numbers that correspond to the set $\left\{1,2,3,..,N\right\} $. \\
To form the diversified initial population, we used a random method to create a feasible sequence and to generate the other individuals of the initial population.\\
To assess the quality of individuals in a population, we have presented three methods to evaluate the studied problem in a given sequence.\\
The literature has several selection techniques such as proportional selection by tournament, by rank, random selection, etc (see Karasakal and Silav \cite{karasakal2016multi}). For our algorithm, we implemented three selection approaches: the aggregative approach, the Pareto one and NSGA-II.\\
The process of crossover between two parents leads to the birth of two children. In this case, an exchange position is randomly determined(see Vallada and Ruiz \cite{Vallada2011}). The first part of the first child is directly obtained from the first parent. The second part is provided by respecting the order of the remaining tasks as they appear in the second parent tasks. The same process is applied to the second child by reversing the parents. For our algorithm, we implemented the one-point crossover, which is a folklore (see Holland \cite{Holland1975}).\\
Several methods of mutation exist in the literature such as the method of permutation, insertion and inversion. In our case, we used the permutation method of swapping two positions of the individual.
\subsection{Aggregative approach}
\label{sec:42}
To adapt our genetic algorithm to the multi-objective case, we constructed an aggregative selection method that consists in generating weights for each sequence of a given population. To calculate the weights, we used an experimental design method called Uniform Design (Leung and Wang \cite{Leung2000}). We choose a new population by a scaling method, which consists in calculating the weighted sum of normalized objective functions. Several combinations of weight are considered for the three objective functions (makespan, total tardiness and completion time). Each combination of these weights transforms the problem into a mono-objective case. Accordingly, the search directions are uniformly dispersed to the Pareto front in the objective space. With multiple fitness functions, we design a selection scheme to maintain the quality and the diversity of the population. This selection scheme consists in applying at every iteration the fitness functions (See Equation 19) and to sort the individuals of the current population in increasing order according to the Scaling method (See Section \ref{sec:422}). The best individuals are selected to form a new population.\\
In what follows, we will describe the Uniform Design method used for calculating the weight and we will give the formula for the scaling method for the selection of a new population.
\subsubsection{Calculation of weight with Uniform Design}
\label{sec:421}
The main objective of the Uniform Design is to sample a small set of points from a given set of points, so that the selected points are uniformly dispersed. This method is a branch of statistics that has been used to calculate the weight. As an illustration of the Uniform Design method, the reader could consult Leung and Wang \cite{Leung2000}.\\
We consider a unit hyper-cube $C$ in a $K$ dimensions space ($K$ is the number of objectives) and $h$  a point in $C$, Where
$h=(h_1,h_2,..,h_K)^T $, such that  $0\leq h_i\leq 1$ $\forall 1\leq i \leq K$.\\
For any item $h$ from the hyper-cube $C$, we can create a hyper-rectangle $R(h)$ between the center $O$ and $h$, with $O=(0,0,..,0)^T $. This hyper-rectangle is described by the following formula:
\begin{eqnarray}
R(h)=\left\{a \in C / a=(a_1,a_2,..,a_K), 0\leq a_i\leq h_i, \forall 1\leq i \leq K\right\} 
\end{eqnarray}
We consider a set of $X$ points from $C$, We can associate with each point $h$, a subset of $X$ points that belongs to the hyper-rectangle $R(h)$. Let $X(h)$ be the cardinality of such a sub set and $X(h)/X$ the fraction of the points included in the hyper-cube $C$ and  $\prod_{i=1}^K h_i$ is the fraction of volume value of the hyper-rectangle $R(h)$. The uniform design is to determine $X$ points in $C$ such that the following discrepancy is minimized.
 \begin{eqnarray}
\underset{h \in C}{Sup}|\frac{X(h)}{X}-\prod_{i=1}^K h_i|
\end{eqnarray}
The authors presented the $X$ points solution calculated using the uniform matrix $U(K,X)\left\{U_{r,i}\right\}_{X*K}$ given by Fang and Li \cite{Fang1994}.
With $U_{r,i}=(r.\sigma^{i-1} mod X)+1$ and $\sigma$ is a parameter that depends on $X$ and $K$.\\
Now, we consider our problem studied, which consists in optimizing three objectives. In our case, we have $K=3$, we take $X=7$, so $\sigma=3$ (see Leung and Wang \cite{Leung2000}). Using the formula $U_{r,i}$ given by Fang and Li\cite{Fang1994}, we get the following uniform matrix:
\begin{center}
$U(3, 7)\left\{U_{r,i}\right\}_{X*K}= \left(
  \begin{array}{ c c c}
     2 & 4 & 3 \\
    3 & 7 & 5 \\
   4 & 3 & 7 \\
    5 & 6 & 2\\
   6 & 2 &4 \\
    7 & 5 & 6 \\
    1 & 1 &1 \\
  \end{array} \right)$ 
  \end{center}
  We consider the weighting vector $W^r=(w_1^r,w_2 ^r,..,w_K ^r)^T$. The components of this vector are calculated by the following formula:
  \begin{equation}
   w_i ^r=\frac{U_{r,i}}{U_{r,1}+U_{r,2}+..+U_{r,K}},
   \forall  1\leq r\leq X, \forall 1\leq i \leq K
  \end{equation} 
\subsubsection{Scaling method}
\label{sec:422}
We use the weight components vector calculated by using the Uniform Design to build a scaling method that allows us to choose a new population by sorting individuals of the current population in ascending order according to the following formula: 
\begin{equation}
H\left(s\right)=w_1^r *C(s)+ w_2^r *TT
(s)+w_3^r *TC(s)
\end{equation}
Such that,
\begin{equation}
C(s)=\left ( \frac{C_{max}\left(s\right)-\underset{x\in P}{min}\left\{C_{max}\left(x\right) \right\}}{\underset{x\in P}{max}\left\{C_{max}\left(x\right) \right\}-\underset{x\in P}{min}\left\{C_{max}\left(x\right) \right\}} \right )
\end{equation}
\begin{equation}
TT(s)=\left ( \frac{T\left(s\right)-\underset{x\in P}{min}\left\{T\left(x\right) \right\}}{\underset{x\in P}{max}\left\{T\left(x\right) \right\}-\underset{x\in P}{min}\left\{T\left(x\right) \right\}} \right )
\end{equation}
\begin{equation}
TC(s)=\left ( \frac{C_{time}\left(s\right)-\underset{x\in P}{min}\left\{C_{time}\left(x\right) \right\}}{\underset{x\in P}{max}\left\{C_{time}\left(x\right) \right\}-\underset{x\in P}{min}\left\{C_{time}\left(x\right) \right\}} \right )
\end{equation}
where $w_i ^r$ are the vector components of the weight described in the above section, $s$ is a feasible solution from population $P$ and $C_{max}(x)$,  $T(x)$, $C_{time}(x)$, are respectively the makespan, the total tardiness and total completion time of a solution $x \in P$.\\
By exploiting the uniform matrix, we obtain seven evaluation functions (fitness). The list of functions is given by the following formula:
\begin{equation}
\left\lbrace
   \begin{array}{l l l}
     fitness_1=\frac{2}{9}*C(s)+ &\frac{4}{9}*TT(s)+ &\frac{3}{9}*TC(s)\\
     fitness_2=\frac{3}{15}*C(s)+&\frac{7}{15}*TT(s)+&\frac{5}{15}*TC(s)\\
     fitness_3=\frac{4}{14}*C(s)+&\frac{3}{14}*TT(s)+&\frac{7}{14}*TC(s)\\
     fitness_4=\frac{5}{13}*C(s)+&\frac{6}{13}*TT(s)+&\frac{2}{13}*TC(s)\\
     fitness_5=\frac{6}{12}*C(s)+&\frac{2}{12}*TT(s)+&\frac{4}{12}*TC(s)\\
     fitness_6=\frac{7}{18}*C(s)+&\frac{5}{18}*TT(s)+&\frac{6}{18}*TC(s)\\
     fitness_7=\frac{1}{3}*C(s)+&\frac{1}{3}*TT(s)+&\frac{1}{3}*TC(s)\\
   \end{array}
   \right.
\end{equation}
Each combination of these weights transforms the problem into a mono-objective case. For each combination, the genetic algorithm is applied simultaneously, the best feasible solutions are selected to form a new population by sorting the individuals of the current population in increasing order of the weighted objective and the population is stored. At the end of this process, such populations are merged and only the non-dominated solutions are kept.
\subsection{Pareto approach}
\label{sec:43}
We adapt classical genetic algorithm for multi-objective case using the Pareto approach Pareto \cite{Pareto1896}. For each generation, we transform the population $P$ by crossing the non-dominated solutions and mutating the dominated solutions. Then, we concatenate the current population $P$ and the new individuals created by crossover and mutation(see Alberto and Mateo\cite{alberto2011crossover}). The new population is then obtained by keeping all non-dominated solutions. In case the number of non-dominated solutions is less than the population size, we complete the remaining population by the best individuals according to three fairly studied criteria: The one-third of the remaining population by the best individuals according to the makespan criterion and the one-third of the remaining population by the best individuals according to the total tardiness criterion. The best individuals according to the total completion time criterion will complement the rest of the population. In the last generation, only non-dominated solutions are kept.
\subsection{NSGA-II algorithm}
\label{sec:44}
The NSGA-II algorithm is based on the following principles Deb et al \cite{Deb2002}:
\begin{itemize}
\item With each generation $g$, merging the population of parents $P_g$ of size $s$ with the population of children $E_g$ of the same size to build a new population $R_g=P_g \cup E_g$ of size $2*s$.
\item Sort the $R_g$ results population according to the non-dominance criterion. This makes it possible to distribute $R_g$ in several fronts $\left( F_1, F_2,...\right)$. The first fronts contain the best individuals.
\item Building the new parent population $P_{g+1}$ by adding the $F_i$ fronts while the size of  $P_{g+1}$ does not exceed $s$. In the case where the size of the new population is less than $s$, the crowding method is applied.
\end{itemize}
The calculation of the crowding distance of an individual is based on the following principles:
\begin{itemize}
\item Repeat these steps for all objectives.
\item Sort the solutions of an objective in ascending order. 
\item Assign infinite distance for the individuals having extreme values (the first and last according to the sorts). 
\item For all other individuals, calculate the normalized difference of the two adjacent solutions. Add the value and calculate the distance of the current individual.
\end{itemize}
After calculating the crowding distance of $i^{th}$ front $F_i$ from $R_g$, the list of $F_i$ solutions must be sorted in a descending order. The best solution is selected by using the crowded comparison-operator $(\prec_n)$; between two different rank solutions, we choose the one with the smallest rank, if they have the same rank we choose the solution that has the greatest crowding distance. 
\section{Numerical results }
\label{Numerical}
\begin{figure}
\centering
\includegraphics[width=9 cm,height=3 cm]{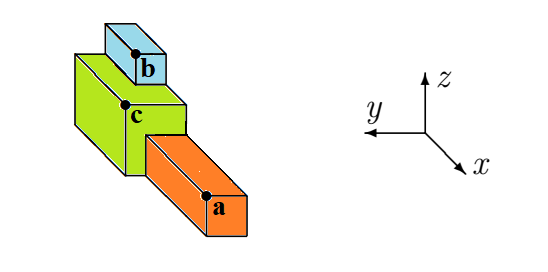}
\caption{Example hypervolume in three dimensions (reproduced from Bradstreet \cite{Bradstreet2011} ).}
\label{fig:21}
\end{figure}

In this section, we present some experimental results obtained on randomly-generated instances. Then, we analyze these results and we provide some conclusions.\\
We implemented our genetic algorithm using a $DEV C++$ compiler on an Intel $Core^{TM}$ i3 4005U CPU 1.7 GHz, 1.7 GHz and 4 GB of RAM.\\
\begin{table}
\caption{Problem types}
\centerline{
\label{tab:3}      
\begin{tabular}{p{2.2cm}p{1.5cm}p{1.5cm}p{1.5cm}p{1.5cm}p{1.5cm}}
\hline\noalign{\smallskip}
$Number\ tasks$ & $Type 1$ & $Type 2$ & $Type 3$& $Type 4$ & $Type 5$  \\
\noalign{\smallskip}\hline\noalign{\smallskip}
$n_1=$ & $n$ &  $n$ &  $n$ & $  n $ & [$n$/2]\\
$n_2=$ & [$n$/2] &  $n$ &  [$n$/2] & $  n $ & [$n$/2]\\
$n_{12}=$ & [$n$/2] &  [$n$/2] &  $n$ & $  n $ & $n$\\
\noalign{\smallskip}\hline
\end{tabular}}
\end{table}
We randomly generated instances by taking into account five types of problems illustrated in Table \ref{tab:3} presented by Manaa and Chu \cite{Manaa2010}. The parameter $n$ is an integer ($n \in \left\{10, 20\right\}$), and [x] corresponds to the integer part of $x$. The variables $n_1$, $n_2$ and $n_{12}$ respectively represent the number of $P_1-tasks$, $P_2-tasks$ and $P_{12}-tasks$.\\
For these five types of problems, Manaa and Chu \cite{Manaa2010} considered the distribution of the three types of tasks and the number of tasks on each processor (load on the processor).\\
For $Type 4$, the distribution of tasks is balanced ($n_1 = n_2 = n_{12}= n$) and the distribution of the load on each processor ($P1: n_1 + n_{12} = 2n$  and $P2: n_2 + n_{12}= 2n$) is therefore balanced.\\
For $Type 1$, the number of tasks $n_1$ exceeds that of the two other types ($n_1> n_2 $ and $n_1> n_{12}$), while  the processor $P1$ is more loaded than $P2$. For $Type 5$, the number of tasks $P_{12}$, which requires the use of the two processors, exceeds that of tasks of the other two types ($n_{12}> n_1$ and $n_1> n_2$). But, the distribution of load on the processors is balanced.\\
For $Type 2$, the load on the processors is balanced, which is not the case for $Type3$. The processing times are randomly generated from the set $\left\{0,..,50\right\}$.\\
The values $r_j$ are randomly generated from the set $\left\{0..L\right\}$, with $L$ equal to the integer part of: $\alpha*(s_1 + s_2+s_{12})$, where $\alpha \in \left \{0.5, 1, 1.5\right\}$ and $s_1, s_2$ and $s_{12}$ are respectively the totals of the processing time of $P_1-tasks$, $P_2-tasks$ and $P_{12}-tasks$.\\
The due dates $d_j$ are randomly generated from the set $\left\{r_j+p_j,..,r_j+p_j+L  \right\}$.\\
We consider that the group of instances represents the set of instances having the same parameters $n$, $\alpha$ and $Type$.\\
For the experimental results, 10 instances of each group are generated and the average values are provided. We fixed the number of generations to $2*Nb$ for each population where $Nb$ is the number of tasks to be processed. Some preliminary tests have motivated our choices.\\
We used two metrics to measure the quality of the Pareto front: the hypervolume indicator (HV) and the number of solutions in the optimal front (ND). The hypervolume is one of the most popular metrics for multi-objective optimisation problems (Bradstreet \cite{Bradstreet2011}). For that, we will use this indicator to measure and compare the performance of aggregative, Pareto and NSGA-II algorithms we propose to solve our studied problem.\\
To calculate the hypervolume of a set of non-dominated points, we used the program implemented by Fonseca et al. \cite{Fonseca2018}. The hypervolume measure uses a reference point (the worst value in each creteria). In this paper, we use the initial solutions calculated by the proposed methods to determine the reference points.\\
To compare the performance of the proposed algorithms, we used the HV ratio (denoted by $HV_r$). 
The following formula compute $HV_r$ the distance between the solutions and the lower bounds:
\begin{equation}
HV_r=\left (1- \frac{HV_{LB}-HV_{Algorithm}}{HV_{LB}} \right )
\end{equation}
Let a(11, 4, 4), b(3, 3, 10) and c(5, 6, 7) be a set of non-dominated points presented in an orthonormal coordinate system. The coordinates (x, y, z) correspond to criterion 1, criterion 2, and criterion 3, respectively. The hypervolume of the set is the volume of the space covered by points a-c-b.
Figure \ref{fig:21} presents an example to illustrate the notion of the hypervolume indicator, for the three objectives.
\\The remainder of this section is organized as follows. In Sect. \ref{sec:41}, we will present  the numerical results in terms of average computation of time. Sect. \ref{sec:42} compares the results of the three proposed approaches.
\subsection{Computation time }
\label{sec:51}
Tables \ref{tab:4} summarizes the numerical results in terms of average computation of time (second). These results show the importance of distinguishing not only the total number of tasks and the length of the release dates interval, but also the different types of problems. The results also show that instances corresponding to the problem of $Type 4$ (with the largest number of tasks compared to other types and the tightest distribution of the release dates with $\alpha= 0.5$) are the most difficult to solve.\\  
\begin{table}
\caption{Computation time for n=10 (in second)}
\centerline{
\label{tab:4}      
\begin{tabular}{p{1.6cm}p{1.6cm}p{1.6cm}p{1.6cm}p{1.6cm}p{1.6cm}}
\hline\noalign{\smallskip}
$n=10$ & $Type 1$ & $Type 2$ & $Type 3$& $Type 4$ & $Type 5$  \\
\noalign{\smallskip}\hline\noalign{\smallskip}
$\alpha=0.5$ &0.086 & 0.120 & 0.188 & $  0.260 $ & $ 0.111$\\
$\alpha=1$ & 0.135 &  0.111 &  0.141 & $  0.134 $ & $0.157$\\
$\alpha=1.5$ & 0.098 &  0.128 &  0.122 & $ 0.130 $ & $0.115$\\
\noalign{\smallskip}\hline
\end{tabular}}
\centerline{\\Note: Aggregative GA results.\\}
\centerline{
\begin{tabular}{p{1.6cm}p{1.6cm}p{1.6cm}p{1.6cm}p{1.6cm}p{1.6cm}}
\hline\noalign{\smallskip}
$n=10$ & $Type 1$ & $Type 2$ & $Type 3$& $Type 4$ & $Type 5$  \\
\noalign{\smallskip}\hline\noalign{\smallskip}
$\alpha=0.5$ & 0.040 &  0.055 & 0.126 & $  0.100 $ & $ 0.058$\\
$\alpha=1$ & 0.060 &  0.059 &  0.078 & $  0.81 $ & $0.091$\\
$\alpha=1.5$ & 0.048 &  0.073 &  0.099 & $ 0.088 $ & $0.084$\\
\noalign{\smallskip}\hline
\end{tabular}}
\centerline{\\Note: Pareto GA results.\\}
\centerline{
\begin{tabular}{p{1.6cm}p{1.6cm}p{1.6cm}p{1.6cm}p{1.6cm}p{1.6cm}}
\hline\noalign{\smallskip}
$n=10$ & $Type 1$ & $Type 2$ & $Type 3$& $Type 4$ & $Type 5$  \\
\noalign{\smallskip}\hline\noalign{\smallskip}
$\alpha=0.5$ & 0.072 &  0.121 & 0.188 & $  0.260 $ & $0.112$\\
$\alpha=1$ & 0.128 &  0.112 &  0.151 & $  0.144 $ & $0.158$\\
$\alpha=1.5$ & 0.104 &  0.129 &  0.122 & $ 0.130 $ & $0.125$\\
\noalign{\smallskip}\hline
\end{tabular}}
\centerline{\\Note: NSGA-II results.\\}
\end{table}
For the aggregative and NSGA-II methods, our genetic algorithm requires an average of computation time equal to $0.260$ seconds for the type of problem $Type 4$ (with $\alpha = 0.5$). For the Pareto method, the average of computation time is equal to $0.100$ seconds. 
The problem of $Type 1$ remains the easiest to solve. The numerical results also reveal that the aggregative approach and NSGA-II require an average of computation time more than the Pareto approach.
\begin{table}[ht]
\caption{Computation time for n=20 (in second)}
\centerline{
\label{tab:5}      
\begin{tabular}{p{1.6cm}p{1.6cm}p{1.6cm}p{1.6cm}p{1.6cm}p{1.6cm}}
\hline\noalign{\smallskip}
$n=20$ & $Type 1$ & $Type 2$ & $Type 3$& $Type 4$ & $Type 5$  \\
\noalign{\smallskip}\hline\noalign{\smallskip}
$\alpha=0.5$ & 0.152 & 0.170 & 0.237 & $ 0.225 $ & $ 0.168$\\
$\alpha=1$ & 0.186 &  0.180 & 0.300 & $  0.278 $ & $0.192$\\
$\alpha=1.5$ & 0.139 &  0.217 & 0.227 & $ 0.285 $ & $0.204$\\
\noalign{\smallskip}\hline
\end{tabular}}
\centerline{\\Note: Aggregative GA results.\\}
\centerline{
\begin{tabular}{p{1.6cm}p{1.6cm}p{1.6cm}p{1.6cm}p{1.6cm}p{1.6cm}}
\hline\noalign{\smallskip}
$n=20$ & $Type 1$ & $Type 2$ & $Type 3$& $Type 4$ & $Type 5$  \\
\noalign{\smallskip}\hline\noalign{\smallskip}
$\alpha=0.5$ & 0.085 &  0.125 & 0.105 & $  0.088 $ & $ 0.086$\\
$\alpha=1$ & 0.071 &  0.120 &  0.228 & $  0.287 $ & $0.240$\\
$\alpha=1.5$ & 0.124 &  0.161 &  0.223 & $ 0.250 $ & $0.168$\\
\noalign{\smallskip}\hline
\end{tabular}}
\centerline{\\Note: Pareto GA results.\\}
\centerline{
\begin{tabular}{p{1.6cm}p{1.6cm}p{1.6cm}p{1.6cm}p{1.6cm}p{1.6cm}}
\hline\noalign{\smallskip}
$n=20$ & $Type 1$ & $Type 2$ & $Type 3$& $Type 4$ & $Type 5$  \\
\noalign{\smallskip}\hline\noalign{\smallskip}
$\alpha=0.5$ & 0.152 &  0.170 & 0.241 & $  0.218 $ & $ 0.172$\\
$\alpha=1$ & 0.186 &  0.178 &  0.281 & $  0.289 $ & $0.186$\\
$\alpha=1.5$ & 0.140 &  0.121 &  0.231 & $ 0.300 $ & $0.192$\\
\noalign{\smallskip}\hline
\end{tabular}}
\centerline{\\Note: NSGA-II results.\\}
\end{table}
\\From Table \ref{tab:5}, our genetic algorithm with NSGA-II approach requires an average computation time equal to $0,300$ seconds for the type of problem $Type 4$ (with $\alpha= 1.5$). In Manaa and Chu \cite{Manaa2010} the branch-and-bound algorithm to minimize the makespan criterion, needs in average more than $574$ seconds to find the optimal solution.
The problem of $Type1$ (having the smallest number of tasks compared to others) requires less computation time compared to other problems. This can be justified by the fact that the processors are loaded with less than the number of bi-processor tasks compared to other cases.
\subsection{Solution quality}
\label{sec:52}
To compare the three objectives of our Aggregative, Pareto and NSGA algorithms, we use the average quality of the three objectives: $C/LBC$, $TT/LBTT$ and $TC/LBTC$. Furthermore, the number of non-dominated solutions and HV ratio will be used to measure the performance of the three proposed algorithms.
\begin{table}[ht]
\caption{Quality of Aggregative GA for n=10}
\centerline{
\label{tab:6a} 
\begin{tabular}{p{1.6cm}p{1.6cm}p{1.6cm}p{1.6cm}p{1.6cm}p{1.6cm}}
\hline\noalign{\smallskip} 
$(Type$, $\alpha)$ & $C/LBC$ & $TT/LBTT$ & $TC/LBTC $& $ND$ & $HV_r(\%)$  \\
\noalign{\smallskip}\hline\noalign{\smallskip}
$\left(1, 0.5\right)$ & 1.111 & 1.277 & 1.410 & 2.5 & 43.479\\
$\left(1, 1\right)$ & 1.124 &  4.542 & 1.212 & 2.8  & 45.390\\
$\left(1, 1.5\right)$ & 1.028 & 8.200 & 1.160 & 2.8 & 78.070\\
$\left(2, 0.5\right)$ & 1.120 & 3.163 & 1.505 & 2.3 & 44.134\\
$\left(2, 1\right)$ & 1.111 & 5.054 & 1.360 & 3.2 & 46.209\\
$\left(2, 1.5\right)$ & 1.089 &  35.000 & 1.350 & 2.9 & 44.287\\
$\left(3, 0.5\right)$ &1.129 & 2.155 & 1.718 & 2.3 & 30.175\\
$\left(3, 1\right)$ & 1.128 & 8.323 & 1.389 & 2.4 & 45.802\\
$\left(3, 1.5\right)$ & 1.103 & 20.222 & 1.341 & 2.6 & 39.948\\
$\left(4, 0.5\right)$ &1.192 & 1.943 & 1.781 & 2.5 & 31.052\\
$\left(4, 1\right)$ & 1.201 & 12.659 &1.466 & 2.6 & 35.217\\
$\left(4, 1.5\right)$ & 1.151 & 15.938 & 1.325 & 2.5 & 24.329\\
$\left(5, 0.5\right)$ & 1.121 & 1.172 & 1.602 & 2.5 & 40.113\\
$\left(5, 1\right)$ & 1.087 & 3.214 & 1.274 & 2.3 & 64.183\\
$\left(5, 1.5\right)$ & 1.031 & 13.889 & 1.198 & 2.8 & 78.984\\
\noalign{\smallskip}\hline
\end{tabular}}
\end{table}
\begin{table}[ht]
\caption{Quality of Pareto GA for n=10}
\centerline{
\label{tab:6b}      
\begin{tabular}{p{1.6cm}p{1.6cm}p{1.6cm}p{1.6cm}p{1.6cm}p{1.6cm}}
\hline\noalign{\smallskip}
$(Type$, $\alpha)$ & $C/LBC$ & $TT/LBTT$ & $TC/LBTC $& $ND$ & $HV_r(\%)$  \\
\noalign{\smallskip}\hline\noalign{\smallskip}
$\left(1, 0.5\right)$ & 1.215 & 1.614 & 1.499 & 2.3 & 14.913\\
$\left(1, 1\right)$ & 1.194 & 2.958 & 1.162 & 2.9 & 26.078\\
$\left(1, 1.5\right)$ & 1.046 & 7.900 & 1.151 & 2.1 & 66.143\\
$\left(2, 0.5\right)$ & 1.326 & 3.050 & 1.582 & 2.2 & 43.716\\
$\left(2, 1\right)$ & 1.243 & 3.000 & 1.298 & 3.3 & 12.818\\
$\left(2, 1.5\right)$ & 1.050 & 15.222 & 1.181 & 3.2 & 65.201\\
$\left(3, 0.5\right)$ &1.257 & 2.521 & 1.808 & 2.9 & 39.814\\
$\left(3, 1\right)$ & 1.217 & 4.677 & 1.328 & 3.3 & 22.822\\
$\left(3, 1.5\right)$ & 1.133 & 6.222 & 1.258 & 3.2 & 29.033\\
$\left(4, 0.5\right)$ & 1.244 & 1.385 & 1.652 & 2.9 & 20.596\\
$\left(4, 1\right)$ & 1.141 & 5.205 & 1.335 & 3.7 & 50.022\\
$\left(4, 1.5\right)$ & 1.137 & 11.469 & 1.297 & 4.1 & 28.614\\
$\left(5, 0.5\right)$ & 1.289 & 1.167 & 1.629 & 3.1 & 24.805\\
$\left(5, 1\right)$ & 1.120 &  2.786 & 1.213 & 4.1 & 52.979\\
$\left(5, 1.5\right)$ & 1.121 &  9.000 & 1.118 & 2.9 & 35.172\\
\noalign{\smallskip}\hline
\end{tabular}}
\end{table}
\begin{table}[ht]
\caption{Quality of NSGA-II for n=10}
\centerline{
\label{tab:6c}      
\begin{tabular}{p{1.6cm}p{1.6cm}p{1.6cm}p{1.6cm}p{1.6cm}p{1.6cm}}
\hline\noalign{\smallskip}
$(Type$, $\alpha)$ & $C/LBC$ & $TT/LBTT$ & $TC/LBTC $& $ND$ & $HV_r(\%)$  \\
\noalign{\smallskip}\hline\noalign{\smallskip}
$\left(1, 0.5\right)$ & 1.181 & 1.422 & 1.449 & 4.3 & 22.261\\
$\left(1, 1\right)$ & 1.209 &  4.929 & 1.188 & 4.7 & 22.822\\
$\left(1, 1.5\right)$ & 1.077 &  7.100 & 1.143 & 2.9 & 47.854\\
$\left(2, 0.5\right)$ & 1.154 & 3.025 & 1.530 & 2.4 & 33.307\\
$\left(2, 1\right)$ & 1.202 &  3.054 & 1.304 & 2.6 & 20.354\\
$\left(2, 1.5\right)$ & 1.058 &  18.222 & 1.204 & 1.9 & 60.583\\
$\left(3, 0.5\right)$ & 1.157 & 2.294 & 1.751 & 3.5 & 21.338\\
$\left(3, 1\right)$ & 1.186 &  4.935 & 1.267 & 3.1 & 29.630\\
$\left(3, 1.5\right)$ & 1.156 & 11.167 & 1.271 & 3.5 & 21.746\\
$\left(4, 0.5\right)$ & 1.225 & 1.930 & 1.700 & 3.7 & 24.070\\
$\left(4, 1\right)$ & 1.208 & 7.114 & 1.330 & 2.2 & 33.697\\
$\left(4, 1.5\right)$ & 1.177 & 13.813 & 1.335 & 2.6 & 17.627\\
$\left(5, 0.5\right)$ & 1.156 & 1.036 & 1.567 & 5.1 & 29.078\\
$\left(5, 1\right)$ & 1.212 &  2.893 & 1.281 & 3.6 & 29.292\\
$\left(5, 1.5\right)$ & 1.148 & 6.444 & 1.162 & 3.6 & 26.050\\
\noalign{\smallskip}\hline
\end{tabular}}
\end{table}
\\Tables \ref{tab:6a}, \ref{tab:6b} and \ref{tab:6c} present the results of the aggregative, Pareto and NSGA-II algorithms for $n=10$. Column 1 indicates the types of problems $(Type$, $\alpha)$ considered in the distribution of the three types of tasks and the number of tasks on each processor. In columns 2, 3 and 4, we present the average quality of the makespan, the total tardiness and the total completion time respectively from instances randomly generated: $(C, TT, TC, ND, HV_r)$ denote respectively the makespan, the total tardiness, the total completion time, the average number of non-dominated solutions and the HV ratio. $(LBC, LBTT, LBTC)$ denote respectively the lower bounds of makespan,  total tardiness and total completion time. Column 5 presents the average number of non-dominated solutions for each problem. Finally, in column 6 we present the results of the HV ratios.\\
The results for the three approaches listed in Tables \ref{tab:6a}, \ref{tab:6b} and \ref{tab:6c} show that the aggregative selection technique is more effective on the makespan criterion for the problems of $Type1$, $Type2$ with($\alpha= 0.5$, $\alpha= 1$), $Type3$ with($\alpha= 0.5$, $\alpha= 1$) and $Type5$.  For the problems of $Type 4$ with ($\alpha= 1$, $\alpha= 1.5$) Pareto is more efficient.\\
The quality of the solutions found by the NSGA-II approach is good on total tardiness criterion for the problem of $Type 1$ with ($\alpha= 0.5$, $\alpha= 1$), $Type3$ with ($\alpha= 0.5$, $\alpha= 1$) and $Type 5$. For the problem of $Type 2$ with ($\alpha= 1.5$, $\alpha= 1$), $Type 3$ with ($\alpha= 1.5$, $\alpha= 1$) and $Type 4$ with ($\alpha= 1.5$, $\alpha= 1$) Pareto is more efficient and the results of aggregative selection technique  are very bad in this cases.\\ 
For the total completion time criterion, the results with NSGA-II algorithm is more efficient. The results with the aggregative approach are close to the lower bounds in some cases and quite far from these lower bounds for the other cases. The results found by the three approaches are close to the lower bounds for the makespan criterion. In some cases, the results of total tardiness are close to the lower bounds with NSGA and Pareto approach.
\begin{figure}
\centering
\includegraphics[width=12 cm,height=5.5 cm]{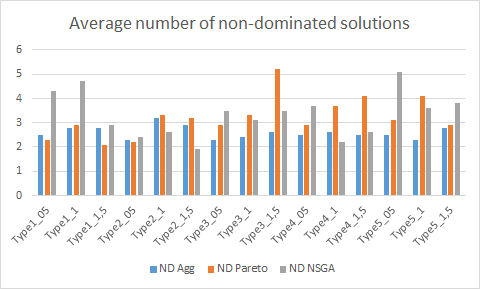}
\caption{Average number of non-dominated solutions for n =10}
\label{fig:3}      
\end{figure}
\\The graphical representation of the front size described in Figure \ref{fig:3} shows that the space of solutions found by NSGA-II and Pareto technique are the most diverse in many cases containing a significant number of non-dominated solutions (between 2 and 5 solutions). This is justified by the fact that this approach ensures elitism by archiving non-dominated solutions in the evolution from one generation to another. The solution space remains the least diversified with the aggregative approach, but it is most effective for the makespan criterion.\\
\begin{figure}
\centering
\includegraphics[width=12 cm,height=5.5 cm]{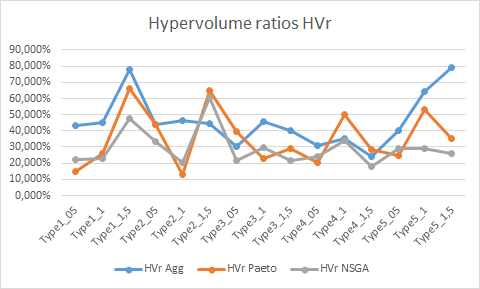}
\caption{Hypervolume ratios for n =10}
\label{fig:4}      
\end{figure}
Figure \ref{fig:4} presents the results of the HV ratios for the three proposed algorithms. The results for the three criteria show that the aggregative selection technique is more effective for problems of $Type 1$ and $Type5$. For the problems of $Type 2$ with ($\alpha= 1.5$) and $Type 4$ with ($\alpha= 1$), the aggregative approach is slightly less efficient.\\
For the Pareto method, the HV ratio is very bad and strongly decreases with the problem of $Type1$ with ($\alpha= 0.5$), $Type2$ with ($\alpha= 1$) and $Type4$ with ($\alpha= 0.5$). The three techniques are almost identical for the problems of $Type4$ with ($\alpha= 1$, $\alpha= 1.5$).\\
Tables \ref{tab:7a}, \ref{tab:7b} and \ref{tab:7c} present the results of the three proposed algorithms for $n=20$.
\begin{table}[ht]
\caption{Quality of Aggregative GA for n=20}
\centerline{
\label{tab:7a}  
\begin{tabular}{p{1.6cm}p{1.6cm}p{1.6cm}p{1.6cm}p{1.6cm}p{1.6cm}}
\hline\noalign{\smallskip} 
$(Type$, $\alpha)$ & $C/LBC$ & $TT/LBTT$ & $TC/LBTC $& $ND$ & $HV_r(\%)$  \\
\noalign{\smallskip}\hline\noalign{\smallskip}
$\left(1, 0.5\right)$ & 1.213 & 4.339 & 1.923 & 2.3 & 21.058\\
$\left(1, 1\right)$ & 1.205 &  15.904 & 1.512 & 1.9  & 31.696\\
$\left(1, 1.5\right)$ & 1.060 & 58.857 & 1.412 & 1.7 & 61.917\\
$\left(2, 0.5\right)$ & 1.298 & 3.328 & 1.984 & 2.5 & 19.100\\
$\left(2, 1\right)$ & 1.210 & 15.962 & 1.630 & 2.2 & 21.514\\
$\left(2, 1.5\right)$ & 1.108 & 99.889 & 1.579 & 2.0 & 38.618\\
$\left(3, 0.5\right)$ &1.263 & 4.803 & 2.204 & 1.4 & 16.517\\
$\left(3, 1\right)$ & 1.302 & 11.643 & 1.703 & 1.8 & 19.300\\
$\left(3, 1.5\right)$ & 1.157 & 86.722 & 1.593 & 2.4 & 28.814\\
$\left(4, 0.5\right)$ & 1.357 & 4.345 & 2.198 & 2.7 & 12.109\\
$\left(4, 1\right)$ & 1.303 & 20.399 &1.823 & 2.6 & 12.395\\
$\left(4, 1.5\right)$ & 1.134 & 45.449 & 1.629 & 2.2 & 35.425\\
$\left(5, 0.5\right)$ & 1.299 & 2.827 & 2.057 & 2.4 & 16.703\\
$\left(5, 1\right)$ & 1.232 & 9.337 & 1.592 & 1.0 & 27.990\\
$\left(5, 1.5\right)$ & 1.132 & 26.362 & 1.519 & 2.3 & 37.857\\
\noalign{\smallskip}\hline
\end{tabular}}
\end{table}
\begin{table}[ht]
\caption{Quality of Pareto GA for n=20}
\centerline{
\label{tab:7b}      
\begin{tabular}{p{1.6cm}p{1.6cm}p{1.6cm}p{1.6cm}p{1.6cm}p{1.6cm}}
\hline\noalign{\smallskip}
$(Type$, $\alpha)$ & $C/LBC$ & $TT/LBTT$ & $TC/LBTC $& $ND$ & $HV_r(\%)$  \\
\noalign{\smallskip}\hline\noalign{\smallskip}
$\left(1, 0.5\right)$ & 1.249 & 3.751 & 1.868 & 3.2 & 14.496\\
$\left(1, 1\right)$ & 1.156 & 7.137 & 1.372 & 2.3 & 43.479\\
$\left(1, 1.5\right)$ & 1.062 & 40.429 & 1.356 & 5.6 & 61.002\\
$\left(2, 0.5\right)$ & 1.292 & 2.331 & 1.890 & 5.3 & 20.002\\
$\left(2, 1\right)$ & 1.189 & 12.278 & 1.512 & 5.2 & 26.078\\
$\left(2, 1.5\right)$ & 1.083 & 68.815 & 1.507 & 4.4 & 49.480\\
$\left(3, 0.5\right)$ & 1.188 & 3.328 & 2.124 & 4.1 & 31.028\\
$\left(3, 1\right)$ & 1.198 & 6.889 & 1.628 & 4.1 & 37.696\\
$\left(3, 1.5\right)$ & 1.188 & 47.417 & 1.538 & 5.9 & 20.810\\
$\left(4, 0.5\right)$ & 1.352 & 3.719 & 2.111 & 1.7 & 12.632\\
$\left(4, 1\right)$ & 1.293 & 19.214 & 1.928 & 3.1 & 13.698\\
$\left(4, 1.5\right)$ & 1.165 & 43.494 & 1.572 & 6.1 & 26.176\\
$\left(5, 0.5\right)$ & 1.296 & 1.994 & 1.912 & 2.2 & 17.147\\
$\left(5, 1\right)$ & 1.111 &  4.251 & 1.398 & 10.2 & 58.132\\
$\left(5, 1.5\right)$ & 1.141 & 11.224 & 1.351 & 6.6 & 35.141\\
\noalign{\smallskip}\hline
\end{tabular}}
\end{table}
\begin{table}[ht]
\caption{Quality of NSGA-II for n=20}
\centerline{
\label{tab:7c}      
\begin{tabular}{p{1.6cm}p{1.6cm}p{1.6cm}p{1.6cm}p{1.6cm}p{1.6cm}}
\hline\noalign{\smallskip}
$(Type$, $\alpha)$ & $C/LBC$ & $TT/LBTT$ & $TC/LBTC $& $ND$ & $HV_r(\%)$  \\
\noalign{\smallskip}\hline\noalign{\smallskip}
$\left(1, 0.5\right)$ & 1.179 & 3.257 & 1.618 & 2.0 & 28.519\\
$\left(1, 1\right)$ & 1.196 & 12.904 & 1.508 & 2.5 & 33.858\\
$\left(1, 1.5\right)$ & 1.089 & 44.762 & 1.400 & 1.8 & 47.563\\
$\left(2, 0.5\right)$ & 1.351 & 2.889 & 1.907 & 2.7 & 12.500\\
$\left(2, 1\right)$ & 1.180 & 14.128 & 1.571 & 2.8 & 28.071\\
$\left(2, 1.5\right)$ & 1.122 & 76.519 & 1.514 & 3.0 & 33.336\\
$\left(3, 0.5\right)$ & 1.242 & 4.342 & 2.162 & 1.8 & 19.925\\
$\left(3, 1\right)$ & 1.346 & 8.609 & 1.597 & 2.0 & 13.705\\
$\left(3, 1.5\right)$ & 1.146 & 66.500 & 1.494 & 1.6 & 32.168\\
$\left(4, 0.5\right)$ & 1.342 & 1.943 & 2.144 & 2.4 & 13.722\\
$\left(4, 1\right)$ & 1.175 & 15.679 & 1.515 & 2.6 & 35.770\\
$\left(4, 1.5\right)$ & 1.192 & 40.831 & 1.571 & 3.2 & 19.474\\
$\left(5, 0.5\right)$ & 1.379 & 2.517 & 2.043 & 2.0 & 7.948\\
$\left(5, 1\right)$ & 1.240 &  6.611 & 1.591 & 2.5 & 26.567\\
$\left(5, 1.5\right)$ & 1.125 & 13.500 & 1.399 & 2.1 & 40.128\\
\noalign{\smallskip}\hline
\end{tabular}}
\end{table}
The results for the makespan criterion described in these tables show that the Pareto approach is more effective for the problems of $Type 3$ with ($\alpha= 0.5$, $\alpha= 1$) and the problems of $Type 5$ with ($\alpha= 1$). For the problem of $Type 3$ with ($\alpha= 1.5$) and $Type 4$ with ($\alpha= 1$) NSGA-II is more effective. For other cases, the three techniques are almost identical. For the five types of problems studied with the makespan criterion, the aggregative approaches and Pareto are identical.\\
The results for the total tardiness criterion described in Tables \ref{tab:7a}, \ref{tab:7b} and \ref{tab:7c} show that the Pareto approach is more effective for the problems of $Type 1$, $Type 3$ with($\alpha= 1.5$)  and $Type 5$ with ($\alpha= 1.5$). For the problem of $Type 2$ with ($\alpha= 0.5$), $Type 3$ with ($\alpha= 0.5$) and $Type 4$ NSGA-II method is the most effective. In other cases, both techniques (NSGA-II and Pareto) are almost identical. The aggregative selection method is less effective for problems of $Type 1$ with ($\alpha= 1.5$), $Type2$ with ($\alpha= 1.5$), $Type3$ with ($\alpha= 1.5$) and $Type5$ with ($\alpha= 1.5$).
\\The numerical results of the total completion time criterion show that the Pareto technique is more effective in many cases ($Type 1$ with ($\alpha= 1$), $Type2$ with ($\alpha= 1$) and $Type5$ ). By looking at these numerical values, we conclude that the three approaches are almost identical and the averages values found are very close. 
\begin{figure}
\centering
\includegraphics[width=12 cm,height=5.5 cm]{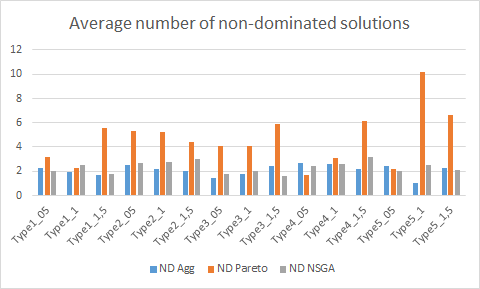}
\caption{Average number of non-dominated solutions for n =20}
\label{fig:5}      
\end{figure}
\\Figure \ref{fig:5} summarizes average number of non-dominated solutions for n =20. The space of solutions found by Pareto technique are the most diverse containing a significant number of non-dominated solutions(between 2 and 10 solutions). The solution space remains the least diversified with the aggregative approach.\\
\begin{figure}
\centering
\includegraphics[width=12 cm,height=5.5 cm]{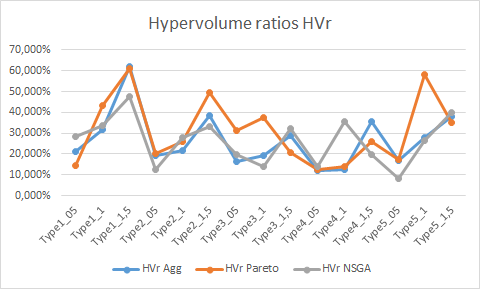}
\caption{Hypervolume ratios for n =20}
\label{fig:6}      
\end{figure}
Figure \ref{fig:6} presents the results of the HV ratios for the three proposed algorithms (for n=20). The results for the three criteria show that the Pareto technique is more effective for problems of $Type 1$ with ($\alpha= 1$),  $Type 2$ with ($\alpha= 1.5$), $Type 3$ with ($\alpha= 0.5$, $\alpha= 1$) and $Type5$ with ($\alpha= 1$. For the problems of $Type 1$ with ($\alpha= 1.5$) and $Type 4$ with ($\alpha= 1.5$), the aggregative approach is more efficient.\\
For the NSGA method, the HV ratio is very bad and strongly decreases with the problem of $Type2$ with ($\alpha= 0.5$) and $Type5$ with ($\alpha= 0.5$). The three techniques are almost identical for the problems of $Type2$ with ($\alpha= 1$) and $Type4$ with ($\alpha= 0.5$).\\
The results found by three approaches listed in the Table \ref{tab:6a}-\ref{tab:7c} are close to the lower bounds for the makespan criterion. In many cases, the results for the total completion time are close to the lower bounds for NSGA-II and Pareto approach. For total tardiness, the results are quite far from lower bounds with the three approaches studied.

\section{Conclusions and perspectives}
\label{conc}
We studied a multi-objective scheduling problem on two dedicated processors to optimize three criteria; the makespan, the total tardiness and the total completion time. In this study, we exploited the lower bound constructed for each criterion, the hypervolume indicator (HV) and the number of solutions in the optimal front (ND) to assess the quality of the solutions found by NSGA-II algorithm, Pareto and aggregative methods proposed for solving the multi-objective problem. To generate the weight for the aggregative approach, we used the method of Uniform Design (proposed by Leung and Wang \cite{Leung2000}) to choose a variety of solutions uniformly dispersed.\\
The results of the studied problems are encouraging and promising for the makespan and  the total completion time criteria. Each studied technique (aggregative, Pareto or NSGA-II) has an advantage compared to the others according one criterion. In some cases, Pareto techniques and NSGA-II are almost identical. Therefore, it is interesting to study other extensions of these problems in a future work, like the study of scheduling problem on parallel processors. For example minimizing the makespan for the problem \(P2|size_j,r_j|C_{max}\) where $size_j$ is the number of processors required by the task $j$. This problem was proved NP-hard in the strong sense in Blazewicz et al. \cite{Blazewicz2002}. Therefore, it is interesting to test the proposed methods for scheduling these problems on parallel processors ( Vallada and Ruiz \cite{Vallada2012}; Venkata et al. \cite{Venkata2018}).\\ \\
\textbf{Compliance with Ethical Standards\\}
\textbf{Conflict of interest } The authors declare that they have no conflict of interest.



\end{document}